\documentclass[aps,twocolumn,pra,floatfix,groupedaddress,nofootinbib,notitlepage,showpacs,superscriptaddress]{revtex4-1}
\usepackage{amsmath}
\usepackage{times,bm,bbm,bbold,graphicx,graphics,amssymb,amsmath,amsfonts,dsfont,hyperref,color}
\usepackage{mathrsfs,cancel}
\usepackage[utf8]{inputenc}
\usepackage{graphicx}
\usepackage{soul,xcolor}
\setstcolor{red}
\definecolor{mycolor}{rgb}{0.1, 0.1, 0.7}
\usepackage{dsfont}
\usepackage{url}
\usepackage{soul}
\usepackage[vlined,ruled]{algorithm2e}

\DeclareFontFamily{OT1}{pzc}{}
\DeclareFontShape{OT1}{pzc}{m}{it}%
{<-> s * [1.25] pzcmi7t}{}
\DeclareMathAlphabet{\mathpzc}{OT1}{pzc}%
{m}{it}
\hypersetup{
	plainpages=true,
	breaklinks=true,
	hypertexnames=false,
	pageanchor=true,
	colorlinks=true,
	linkcolor={mycolor},
	citecolor={mycolor},
	urlcolor={mycolor},
	anchorcolor={black}
}

\begin{document}
	\title{Qubit readout and quantum sensing with  pulses of quantum radiation}
	\author{Maryam Khanahmadi}
	\email{m.khanahmadi@chalmers.se}
	\affiliation{Department of Microtechnology and Nanoscience, Chalmers University of Technology, 412 96 Gothenburg, Sweden}
	\affiliation{Department of Physics, Institute for Advanced Studies in Basic Sciences, Zanjan 45137, Iran}

	\author{Klaus M{\o}lmer}
	\email{klaus.molmer@nbi.ku.dk}
	\affiliation{Niels Bohr Institute, University of Copenhagen, Blegdamsvej 17, DK-2100 Copenhagen,Denmark } 

	\begin{abstract}
	Different hypotheses about a quantum system such as the logical state of a qubit or the value of physical interaction parameters can be investigated by the interaction with a probe field. Such fields may be prepared in particularly sensitive quantum states, and we can use quantum trajectories to model the stochastic measurement record and conditional evolution of the state of the quantum system subject to its interaction with a travelling pulse of radiation.  Our analysis applies to different measurement strategies and to arbitrary input quantum states of the probe field pulse and it thus permits direct comparison of their metrological advantages. 	\end{abstract}
	\date{\today}
	\maketitle

\section{Introduction}
The motivation in quantum optics to study a variety of so-called non-classical states of light, such as number states, squeezed states, entangled states and Sch\"odinger cat states  has been associated with their use in precision measurement protocols. Very sensitive measurements may thus benefit from the use of probe fields that are prepared in states that change maximally upon the interaction with the object or phenomenon under investigation while displaying minimal variance of the  observable measured \cite{caves1981quantum,lioyd-metrology}.
More general approaches adopt advanced analyses of the information that can be extracted by optimal general measurements on the quantum state \cite{braunstein1994statistical}.

\begin{figure}[ht!]
		\includegraphics[scale=.45]{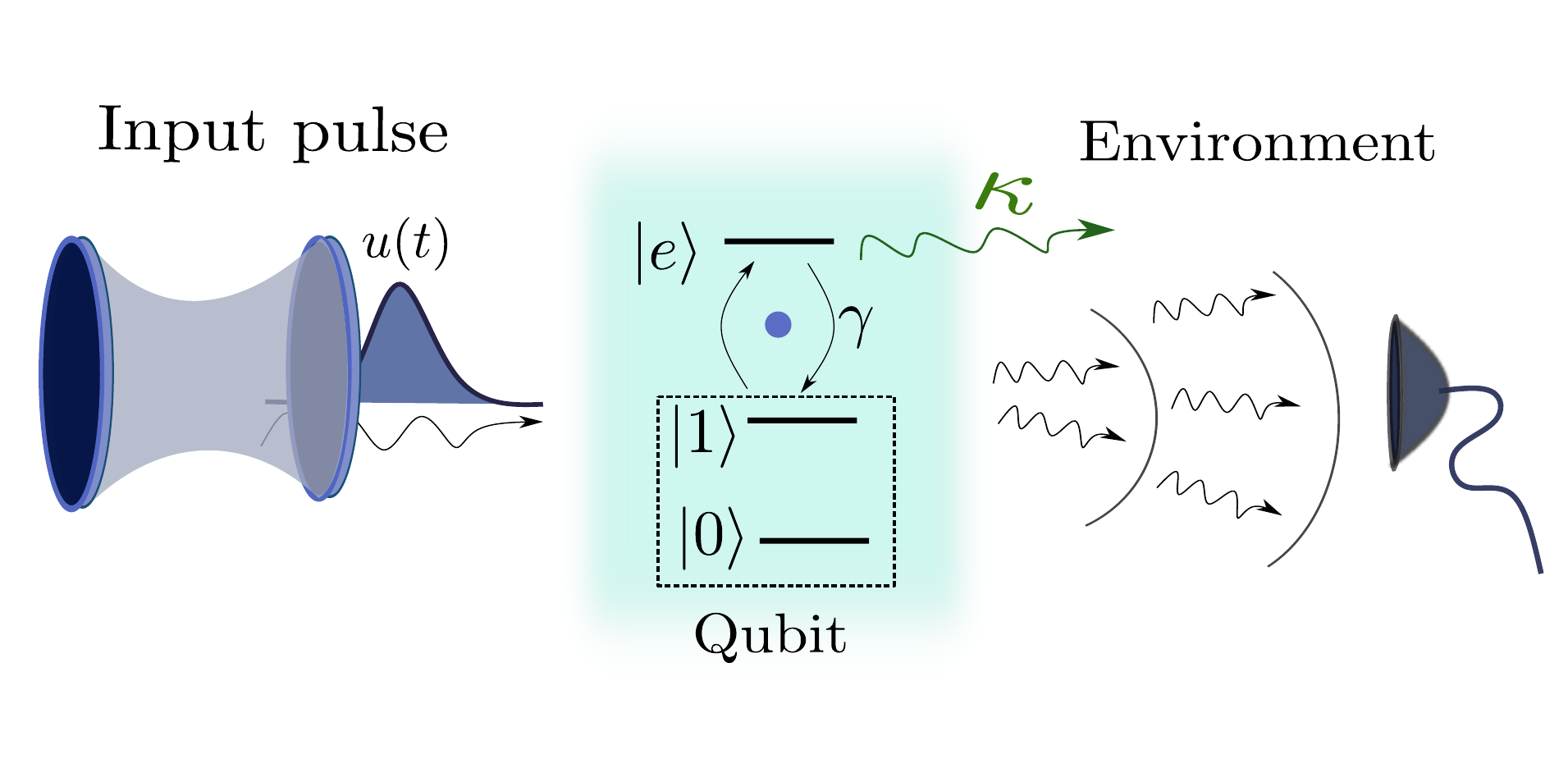}
		\caption{Schematic of qubit state readout by an incoming quantum pulse. The incident light pulse with shape $u(t)$ excites a closed optical transition between the qubit state $|1\rangle$ and excited state $|e\rangle$. The amount of absorption from the pulse and the field correlations induced by the interaction with the qubit system are registered by continuous photon counting or homodyne detection. By introduction of a virtual cavity as the source of the incident pulse, we obtain an effective single mode open systems treatment of the physical interactions.
		}
		\label{F1}
\end{figure}

In this article we consider probing of a quantum system by its interaction with an itinerant travelling wave packet of quantum light or microwave photons, see Fig. \ref{F1}. Propagating quantum states have been proposed to mediate quantum state transfer and quantum interactions between stationary physical systems  \cite{pfaff2017,kimble2008,prlcirac,matsukevich,stute}, but a practical theory for how  a single-mode input pulse of quantum radiation interacts with a local quantum system has only been presented recently \cite{PhysRevA.86.013811,PhysRevA.96.023819,PhysRevLett.123.123604,kiilerich2020quantum,fischer2018scattering}. The purpose of the present study is two-fold: On the one hand, we shall extend previous, simplified treatments and provide a description of how the travelling quantum pulse interacts with matter in a fully time dependent manner, and, on the other hand, we shall present an analysis of the full measurement record from the continuous probing of the  radiation field after the interaction with the system of interest. 

A non-linear scatterer generally produces a multi-mode output field which does not have a manageable quantum state description in terms of a state vector or density matrix. But, in \cite{PhysRevLett.123.123604,kiilerich2020quantum}, it was shown that it is possible to calculate the dynamics of the scatterer by treating the output field as loss in a cascaded systems master equation. 
We shall incorporate the effect of  measurements into this theory by a stochastic unravelling of the master equation. The resulting equation, in turn, forms the basis for a quantum filter theory along the lines of \cite{belavkin1995quantum}. The application of the filter approach was so far restricted to systems probed or excited by classical fields, see, e.g., \cite{gambetta2001state,chase2009single,tsang2012continuous,gammelmark2013bayesian}. Our stochastic cascaded master equation yields results that are  equivalent to the ones obtained  in \cite{PhysRevA.96.023819} by an alternative method. We believe that the derivation and application of our master stochastic equation is more straightforward, and our theory is also more readily  extended and applied for parameter estimation and hypothesis testing.

We present simple examples of our formalism to the readout of the state of a qubit and to the discrimination between discrete values of its physical interaction parameters. We consider photon counting and homodyne detection of the transmitted field, and we note that the theory readily applies also to the precision measurement of continuous parameters cf.,  \cite{gammelmark2013bayesian,gammelmark2014fisher}




The article is structured as follows, in Sec. \ref{theory} we introduce the description of the interaction between a quantum pulse and a discrete quantum system. In Sec. \ref{QT}, We model the stochastic dynamical equation of the quantum pulse and the quantum system due to the back action of the continuous counting or homodyne measurements. In Sec. \ref{read} we present the Baysian inference from the entire signal record about the initial state of a qubit quantum system or a system parameter. We summarize the results and some theoretical considerations on the  applied method in Sec. \ref{SUM}.

\section{Cascaded master equation}\label{theory}

There is a fundamental difference between the time dependent interaction of a quantum system traversing a  field eigenmode confined in a cavity, and the time dependent interaction of an incident single mode pulse of travelling radiation with a localized quantum system. In the former case, the field is restricted to discrete eigenmodes, and the single mode Jaynes-Cummings model may apply to an excellent approximation, while in the latter case, the field is free to explore a continuum of propagating modes, and any non-linearity in the quantum system thus leads to population of a multi-mode output field. Quadratic interaction, leading to linear equations for the field amplitudes, have been dealt by input-output theory \cite{gardiner1985input}, while  scattering of single and two photon wave packets on a two-level system has been solved by scattering theory, \cite{shen2007strongly,witthaut2012photon,mahmoodian2018strongly,yang2022deterministic}. But, outside these exceptions, treatments of the interactions between a quantum pulse and a scatterer seem prohibitively complicated due to the dimensionality of the multi-photon and multi-mode Hilbert space. It has been shown, however, that the theory of cascaded input-output quantum systems\cite{carmichael1993quantum,gardiner1993driving} permits treating the scatterer and a single mode of the radiation field as an open quantum system. This idea was first implemented in \cite{PhysRevA.86.013811,PhysRevA.96.023819}, where effective coupled master equations of the scatterer were associated with each Fock state of the pulse. In this article, we shall adopt a simpler treatment with a virtual cavity that leaks the pulse towards the target, as described by a cascaded master equation  \cite{PhysRevLett.123.123604,kiilerich2020quantum}.
Fig. \ref{F1} shows a system with two stable levels $(|0\rangle,|1\rangle)$ and a state $|e\rangle$ that is excited from $|1\rangle$ by the interaction with the incoming pulse with strength $\sqrt{\gamma}$, and decays spontaneously with rate $\gamma$. We assume for simplicity that the input and output fields are transversally single mode while populating a continuum of  radiation modes propagating from left to right in the figure (chiral coupling).

To obtain a travelling wave packet $u(t)$ as the output from a one-sided quantum cavity with a single mode annihilation operator $a_u$, we must assume a time-dependent coupling strength $g(t)$ between the cavity and the input continuum field $b_{\mathrm{in}}(t)$  
\begin{align}\label{H1}
H_{u,\mathrm{cavity}} = i[g^*(t)b_{\mathrm{in}}^{\dagger}(t)a_u-g(t)b_{\mathrm{in}}(t)a_u^{\dagger}].
\end{align} 
Assuming
\begin{align}\label{T3}
    g(t) = \frac{u^{*}(t)}{\sqrt{1-\int_{0}^{t}\mathrm{d}t'|u(t')|^2}}.
\end{align}
the initial quantum state of the cavity mode is then eventually emitted as a single mode pulse with the pulse shape $u(t)$ \cite{PhysRevLett.123.123604}. 

We treat the quantum scatterer by a Hamiltonian $H_s$ and its exchange of quanta with the incident pulse can now be equivalently described as a coupling to the $u$-cavity mode with the interaction Hamiltonian
\begin{align}\label{H2}
H_{us}=i\frac{\sqrt{\gamma}}{2}[g_u(t)a_u^{\dagger}c - g_u^*(t)a_u c^{\dagger}] 
\end{align}
where $c$ and $c^\dagger$ are raising and lowering operators of the scatterer by the absorption or emission of a quantum of radiation (in our example, $c=|1\rangle\langle e|$).
Radiation propagating to the right of the emitter is composed of the incident pulse (loss by the cavity) and the emission by the quantum scatterer, and is described by the coherent lowering operator on these two systems.
\begin{align}\label{LL}
     L_{0}(t)= g_u^*(t)a_u+\sqrt{\gamma}\sigma.
\end{align}

Finally, the quantum state of the cavity and the system is described by the density matrix $\varrho _{\mathrm{us}}(t)$ which solves the master equation 
\begin{align}\label{I1}
\frac{d\varrho_{us}(t)}{dt} = -i[H,\varrho_{us}(t)]+\sum_{i=0}^{n}\mathcal{D}[L_i]\varrho_{us}(t)
\end{align}
where $H = H_s+H_{us}$, and  the Lindblad terms $\mathcal{D}[L_i]\rho = -\frac{1}{2}(L_i^{\dagger} L_i \rho + \rho L_i^{\dagger} L_i) +  L_i \rho L_i^{\dagger}$ apply for both $L_0$ in \eqref{LL} and for any additional dissipative local Lindblad operators, $L_i, i=1,...,n$. 

The interference of the two lowering operator terms in the Lindblad operator $L_0$ is responsible for the cascaded nature of the master equation: Both $D[L_0]\rho$ and the commutator with the interaction Hamiltonian contribute terms of the form $a_u^{\dagger}c$ ($a_u c^{\dagger}$), multiplying $\rho_{us}$ from the left (right) in Eq.\eqref{I1},  but these terms cancel each other and, effectively,  no atomic excitation is returned to the source cavity mode. 

An exemplary further dissipation mechanism is atomic decay with rate $\kappa$ by emission into a different, unobserved direction, see Fig. \ref{F1}. Such decay is merely represented by one of the  separate Lindblad  terms in Eq.\eqref{I1}, with  
\begin{align} \label{L1-loss}
L_1 = \sqrt{\kappa}|1\rangle\langle e|.
\end{align}


\section{Quantum trajectories for the observed system}\label{QT}

We model the conditional dynamics of the system due to measurement of the transmitted field by replacing the deterministic master equation \eqref{I1} by a corresponding stochastic master equation \cite{wiseman2009quantum,jacobs}. For classical probe fields (coherent states), the cascaded master equation reduces to an equation only for the driven scatterer \cite{paris2017free,kiilerich2020quantum}, while for quantum states of the probe we must retain the stochastic evolution of the combined virtual cavity and scatterer system.        




\subsection{Counting measurement}
 We consider first photon counting measurements on the pulse transmitted by the quantum system; Fig. \ref{F1}.
The stochastic evolution of the density matrix includes a continuous no-jump part for the (un-normalized) density matrix $\Tilde\varrho(t)$,
	\begin{align}\label{C2}
	\mathrm{d}\Tilde{\varrho}(t) = \big[ -i[H,\Tilde{\varrho}(t)]
	+\sum_{i=1}^{n}\mathcal{D}[L_i]\Tilde\varrho(t)
	-\frac{1}{2} \{L_0^{\dagger}L_0,\Tilde{\varrho}(t)\}  \big]\mathrm{d}t
	\end{align}
and the occasional quantum jump part,
\begin{align} \label{eq:jump}
\Tilde{\rho} \rightarrow L_0\Tilde{\varrho}(t)L_0^{\dagger}.
\end{align}
occurring with the probability 
$\delta p =  \langle L_{0}(t)^{\dagger}L_{0}(t) \rangle \mathrm{d}t$. 
Note that operator averages (and the jump probability) can be obtained only after proper normalization of $\Tilde{\varrho}$.
	
\begin{figure}[ht!]
	  \centering
		\includegraphics[scale=.35]{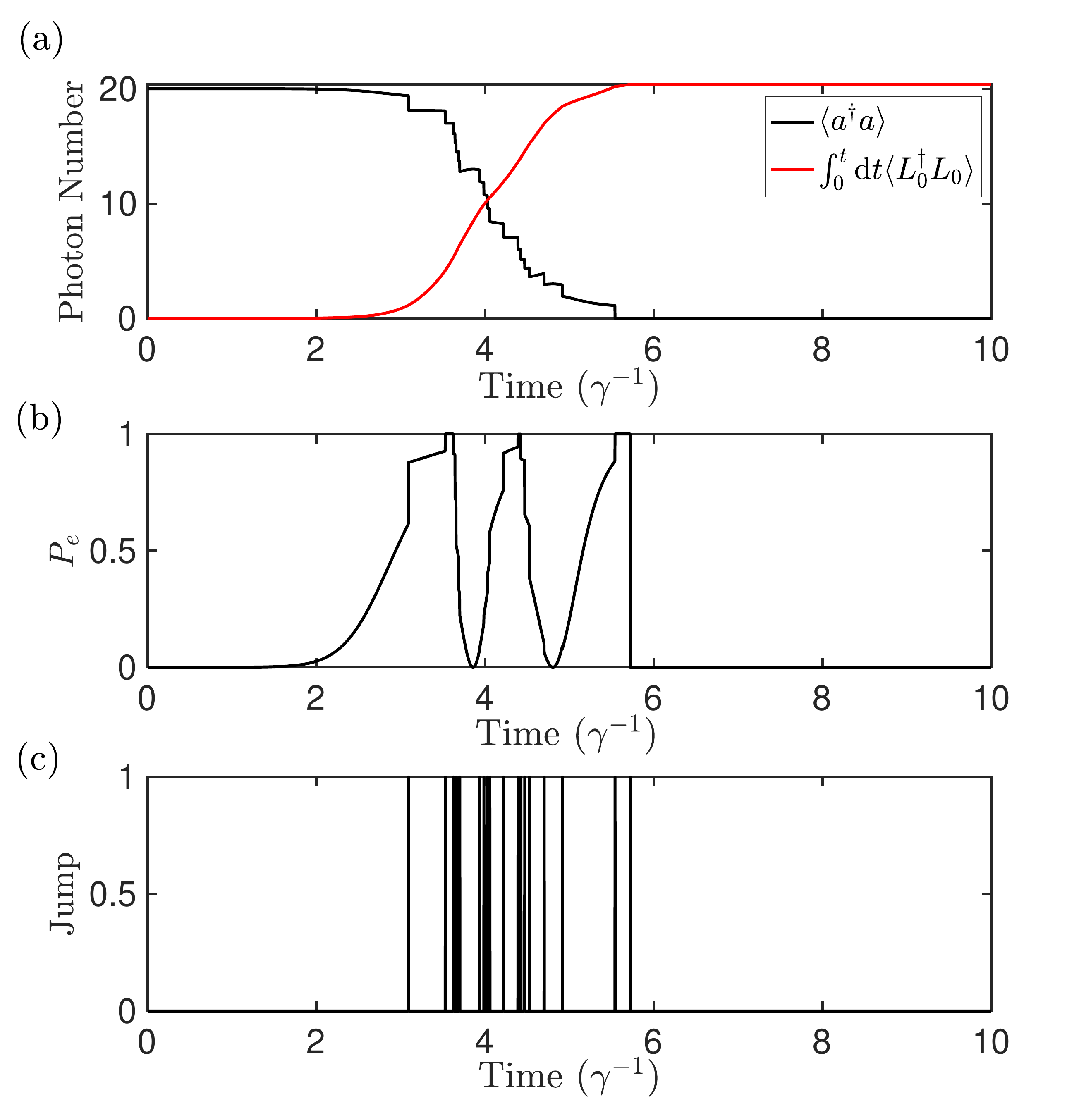}
		\caption{Single trajectory of an atom initialized in the ground state $|\psi\rangle = |1\rangle$ interacting with an incoming $n=20$ photon number state pulse, where the output field is subject to unit efficiency photon counting. We assume no further dissipation channels and that all emission is in the forward direction, cf. Fig.\ref{F1}. Panel (a) shows the mean number of photons in the cavity mode representing the incident Gaussian wave packet (black line) and the integral of the mean photon detection rate (red line). Panel (b), shows the Rabi oscillation-like evolution of the excited state probability $P_e$ of the two level system. Panel (c) shows the (20) individual photon detection events  simulated to occur during the passage of the pulse. }
	\label{f6}
	\end{figure}
\subsection{Homodyne measurement}
In homodyne detection, the radiation is mixed on a beam splitter with a local oscillator with the same frequency. The difference between the photon flux at the two output-ports yields a noisy and continuous signal  $\mathrm{d}Y_t$, with a component proportional to, say, the first quadrature of the quantized field represented by $L_0$, and a white noise term $dW_t$, 
\begin{align}\label{dY}
    \mathrm{d}Y_t = \mathrm{Tr}(L_0\varrho +\varrho L_0^{\dagger})\mathrm{d}t + \mathrm{d}W_t. 
\end{align}
$\mathrm{d}W_t$ is a Wiener increment with zero mean and variance $\mathrm{d}t$. Note that the mean value of the field quadrature is calculated according to the normalized density matrix $\varrho$. 
	\begin{figure}[ht!]
	\centering
		\includegraphics[scale=.41]{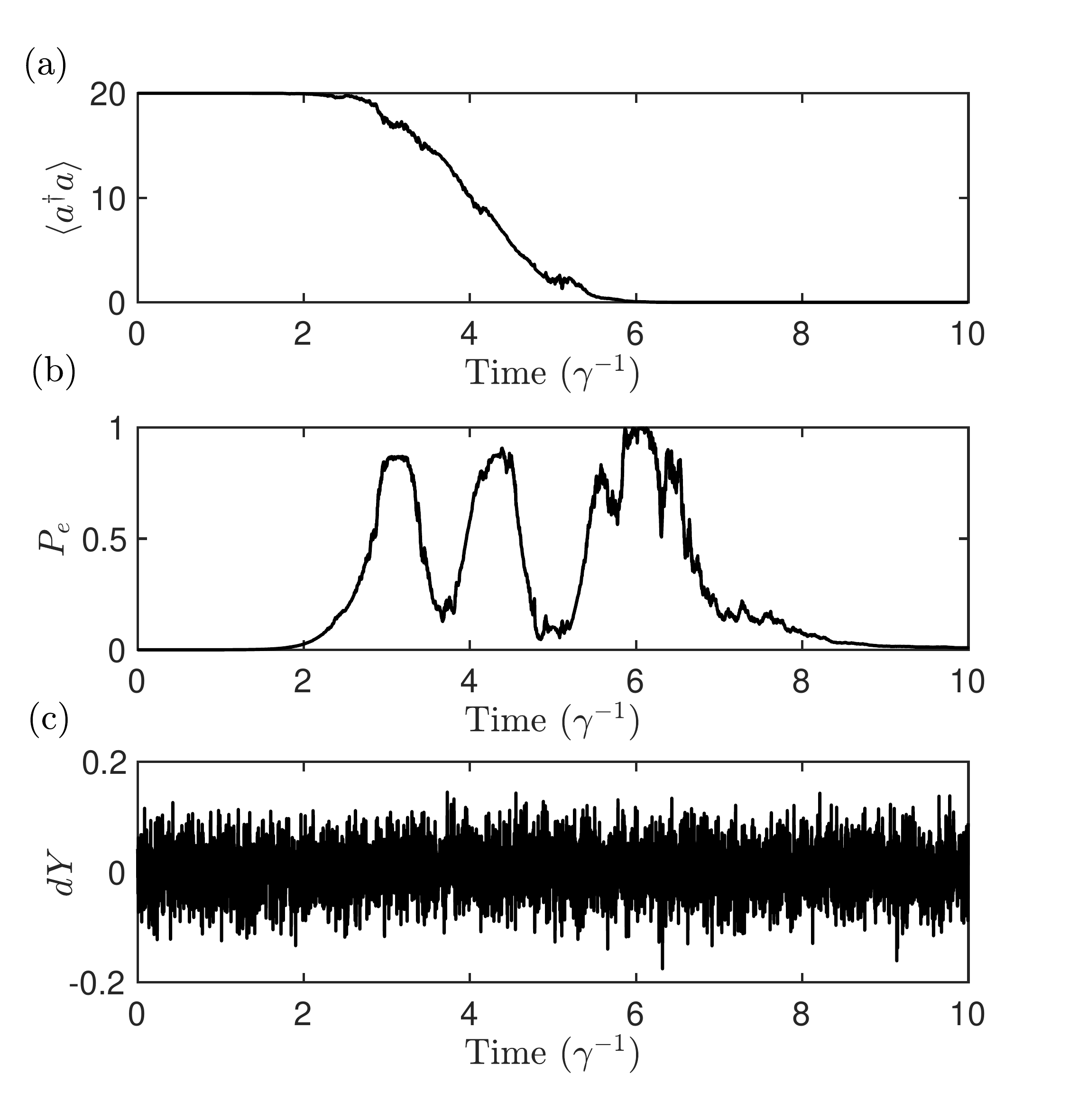}
		\caption{Single trajectory of an atom initialized in the ground state $|\psi\rangle = |1\rangle$ interacting with an incoming $n=20$ photon number state pulse, where the output field is subject to homodyne detection. We assume no further dissipation channels and that all emission is in the forward direction, cf. Fig.1. Panel (a) shows the mean number of photons in the cavity representing the input pulse. Panel (b) shows  the excited state population $P_e$  of the atom. Panel (c) shows the stochastic signal \eqref{dY} obtained during the time interval $[0,10]$.
		}
		\label{f7}
	\end{figure}
	
The measurement back action of the homodyne detection can be incorporated with the deterministic evolution in the un-normalized stochastic differential equation 	
	\begin{align}\label{C3}
	\mathrm{d}\Tilde{\varrho}(t) =& \big[ -i[H,\Tilde{\varrho}(t)]+\sum_{i=0}^{n}\mathcal{D}[L_i]\Tilde{\varrho}(t)
	\big]\mathrm{d}t\nonumber\\
	&+\big[L_0\Tilde{\varrho} + \Tilde{\varrho} L_0^{\dagger}\big] \mathrm{d}Y_t
	\end{align}
A realization of Eq. \eqref{C3} is shown in Fig. \ref{f7}.

\section{Hypothesis testing}\label{read}
If a system is subject to dynamics according to different candidate physical parameters, we may treat those as a set of hypotheses, $\lbrace h_i, i=1,...,m\rbrace$, with corresponding prior probabilities $p_0(h_i)$. Given the outcome of measurements, we can use Bayes rule to update the probabilities and infer the most likely hypothesis. In practice, the stochastic master equation is solved in parallel for density matrices  $\Tilde{\rho}_i(t)$ for each candidate hypothesis $i$. These density matrices  directly constitute a Bayesian filter: the probability for a measurement outcome is, for each different hypotheses $i$, given by $\Tilde{\varrho}_i$ via Born's rule. In fact, the relative probability for a whole data record is accumulated in the norm of the un-normalized density matrix, $p(D|h_i) \propto \mathrm{Tr}(\Tilde{\varrho}_i)$ subject to the stochastic master equation. 

According to Bayes rule, it thus follows that the likelihood of the different hypotheses are updated as $p(h_i|D)\propto p(D|h_i)p_0(h_i)$. Henceforth we shall write $p_i = p(h_i|D)$ for the normalized likelihood, and we note that the Bayesian update may be obtained iteratively over time, cf., the time evolution of the conditional density matrices. The identification of the stochastic master equation with the  Bayesian filter \cite{gambetta2001state,chase2009single,tsang2012continuous,gammelmark2013bayesian}
applies straightfowardly to our quantum pulse master equation, and we are hence able to  directly assess how the probing with quantum pulses can be employed for different sensing tasks.

 \subsection{Qubit state readout}
 
 In the following, we shall consider the determination of the  state of a qubit in the $\{|0\rangle,|1\rangle\}$ subspace of the atomic system depicted in Fig. \ref{F1}. We assume that the incident light pulse interacts with the atom on the closed optical transition  $|1\rangle \leftrightarrow |e\rangle$, and we shall show that this interaction is revealed in the noisy signal records. 
 
 We thus deal with two hypotheses, namely the two possible initial qubit states $|0\rangle$ and $|1\rangle$, and we assume equal initial  probabilities, $p_0(t=0)=p_1(t=0)=1/2$. The best estimate of the actual initial qubit state at any time during the measurements is the one assigned the highest conditional probability $\mathrm{max}(p_0(t),p_1(t))$. That choice, however, will be in error with the remaining probability
 $Q_e = 1 - \mathrm{max}(p_0(t),p_1(t)      )$.

\begin{figure}[ht!]
	\centering
		\includegraphics[scale=.37]{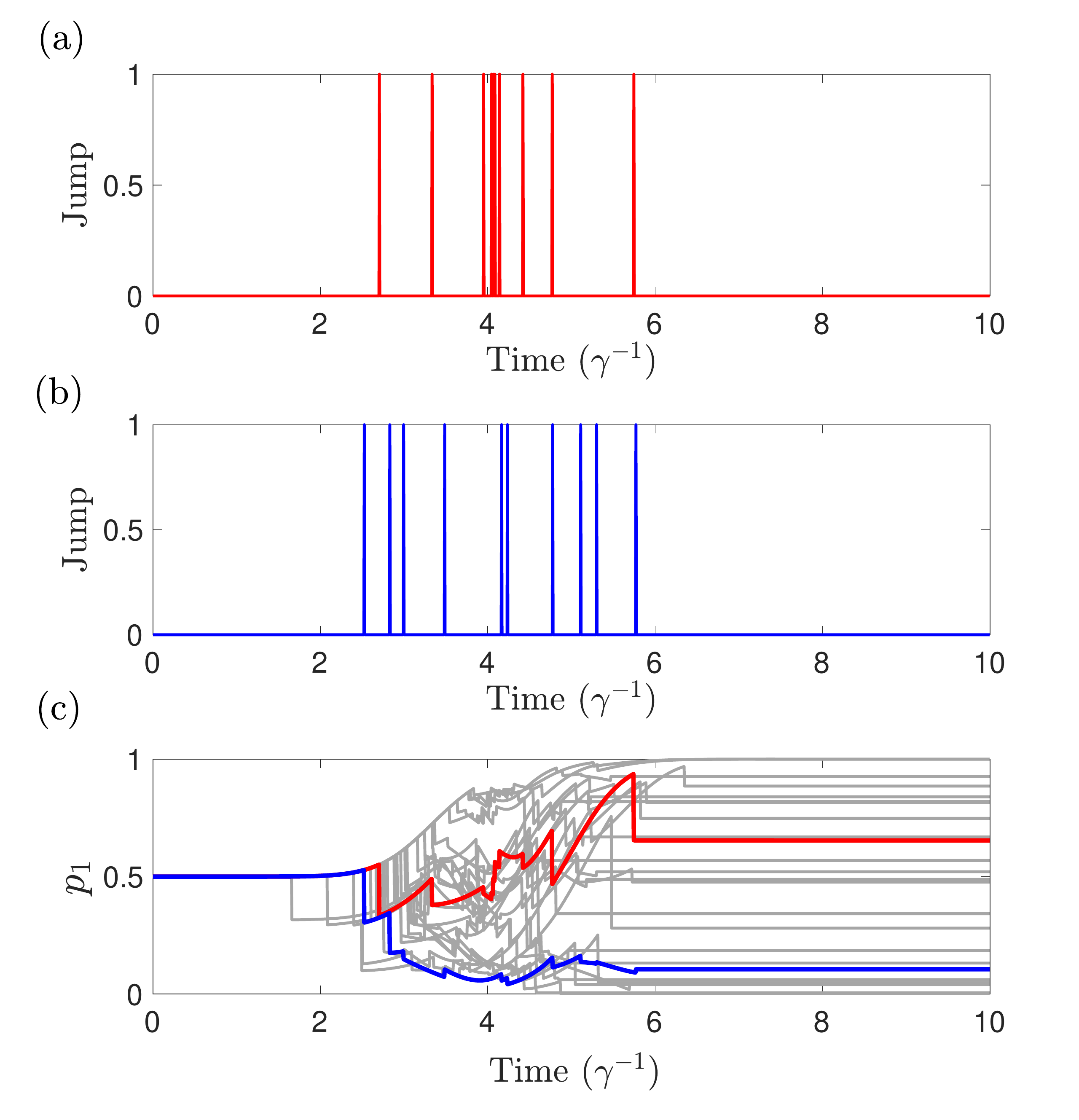}
		\caption{
		 Panels (a) and (b) show simulated photon counting signals for a Gaussian pulse prepared in an $|N=10\rangle$ Fock state, after its interaction with a qubit system prepared in state $|1\rangle$ and $|0\rangle$, respectively. Panel (c) shows the inferred conditional probabilty that the initial state was $|1\rangle$. The red (upper) and blue (lower) solid curves correspond to the simulated detection record in panels (a,b), and the grey thin curves are obtained with other detection records, assuming the initial state $|1\rangle$ and $|0\rangle$ with equal probability. 
		}
		\label{f2}
	\end{figure}

Since all incident photons are eventually detected in the output, the total photon count is independent of the state of the qubit, but the temporal distribution of detector clicks, and the  correlations in the full detection record may still reveal the interaction or not with an effective two-level transition. While these may involve multi-time correlations of a very complex character and, the strength of the Bayesian quantum trajectory analysis is that it requires no formal analysis of such correlations. The quantum trajectory itself constitutes a filter that extracts the maximum information from all available data in the  detection record.  Our analysis is readily applicable with more complex investigations, e.g., of interferometric setups and their exploration with general quantum states of light
 \cite{holland1993interferometric,humphreys2013quantum}.   

\subsection{Results}\label{general}       

Fig.\ref{f2} shows the outcome of different realizations of the photon counting record for a Gaussian wave packet with 10 photons interrogating the three level atom depicted in Fig.\ref{F1}. Sample  detection records are shown in panel (a) and panel (b) with initial qubit states $|1\rangle$ and $|0\rangle$, respectively. Panel (c) shows with solid red and blue curves the inferred, conditional probabilities that the initial state was $|1\rangle$. We see that despite the similarity of the detection records in the upper panels, the inference clearly favors the correct, different initial qubit states. While the total number of detected photons is the same for both hypotheses, their temporal distributions differ as the interaction with the atom can both change the mean intensity profile and the intensity correlations within the pulse. The thin light curves in panel (c) show how the performance of the qubit readout varies for an ensemble of simulated detection records, chosen with random initial qubit states.            

\begin{figure}
    \centering
    \includegraphics[scale=.36]{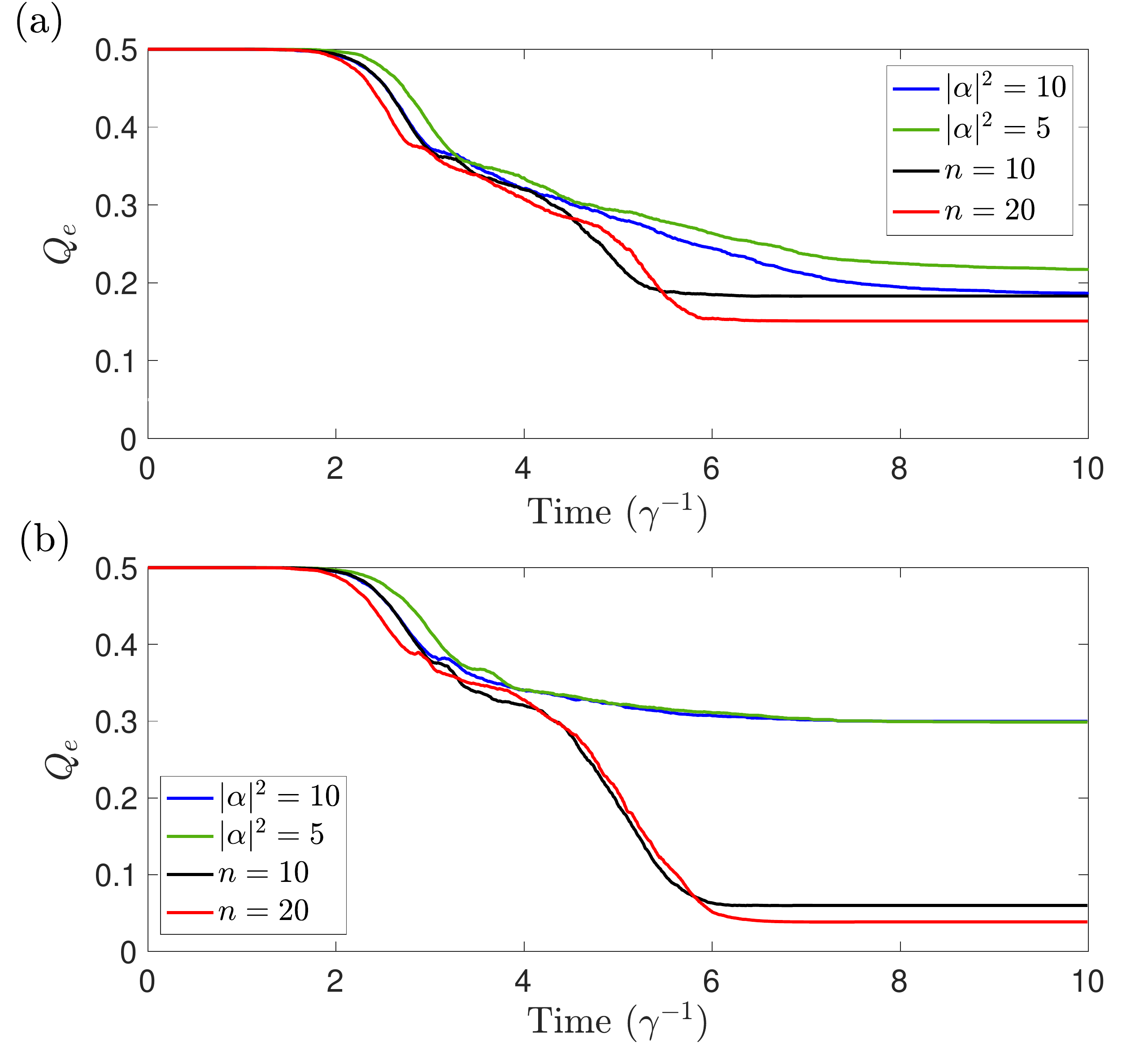}
    \caption{Comparison of the mean error of the qubit readout with Fock and coherent probe pulses of light. The upper panel (a)  shows the results where the atom only emits in the forward direction while panel (b) shows the results when the excited atoms is subject to an extra decay channel with rate $\kappa=\gamma$. In both panels, the green and dark blue curves represent the coherent state  with $\alpha=\sqrt{5}$ and $\sqrt{10}$,  while the black and red curves represent Fock states with 10 and 20 photons, respectively. The results are obtained as averages over 1000 simulated trajectories.}
  \label{F3}
\end{figure}

In Fig. \ref{F3}, we show the average outcome over many simulations of the photon counting detection record. To compare how well different input probe states serve to distinguish the qubit state of the atomic scatterer, the figure shows results for  Fock states with $N=10$ and $20$ photons as well as coherent states $|\alpha\rangle$ with $\alpha =\sqrt{5}$ and $\sqrt{10}$. In the upper panel (a) we assume that the atom only emits radiation in the forward direction, while in panel (b) 
we assume an extra atomic loss process with rate $\kappa=\gamma$. This extra loss is only incurred, if the qubit is in state $|1\rangle$ and hence the detection of a statistically significant reduction in photon number reveals the initial qubit state.

The figures shows the error probability $Q_e$, averaged over 1000 simulated trajectories.  While one may have expected that a larger number of probe photons would have improved the sensing capability significantly, this is seen to play a less important role. The oscillatory structures can be ascribed to the different Rabi oscillation patterns, but the saturation  of the atomic transition reduces their effect on the transmitted signal. Notably, the Fock states yield better inference, and their error probability drops compared to the coherent states at the end of the pulse. We ascribe this to the knowledge of the total number of counts and hence the certain knowledge that excitation resides (or not) in the atom beyond the end of the incident Fock state pulses. In panel (b), we show results for the case where the atom  has an extra loss channel (if excited from state $|1\rangle$). In this case, a  deviation between the total photon count and the incident number of photons points clearly to the initial qubit state $|1\rangle$. For an incident Fock state the assignment error is accordingly smaller than in panel (a) of the figure. For coherent states, photon loss is not discernible with certainty, and we see no similar benefit. Instead, we observe an increased inference error, which may be due to the loss of optical signal and resulting degradation of photon coincidences.

\section{Summary and discussion}
	\label{SUM}
In this article we have presented a general formalism for the interrogation of a quantum system by a quantum probe pulse of radiation. We established the stochastic master equation, which provides the time dependent state of the probe and the scatterer conditioned on the measurement record (assuming counting or homodyne detection). The master equation readily translates to a filter equation providing conditional probabilities for different hypotheses governing the initial state or parameters in the evolution of the system. We illustrated the stochastic formalism by  simulating the detector signal following the interaction between an incoming wavepacket and a qubit system, for which our Bayesian filter exhausts the information available in the detection record about the initial qubit state. Our analysis permits the first quantitative assessment of probing of quantum systems with non-classical pulses of light, and our simulations confirmed the expectation that Fock probe states may be superior over coherent states for such tasks. Our method readily applies for a much wider class of quantum states and measurement settings, i.e., pulses incident on interferometer set-ups.
	
We characterized the achievements of the sensing schemes by averaging the probability of error over many simulated trajectories. Such analyses may be supplemented by extended master equation methods like the protocols developed to calculate the (classical) Fisher information for continuous measurement records in  \cite{gammelmark2013bayesian}.
Following \cite{gammelmark2014fisher,molmer2015hypothesis}, it is also possible to apply extended master equation approaches to provide the quantum Fisher information, which provides a theoretical lower limit to the mean estimation error, or the minimum error probability $Q_e$ obtained by {\it any} hypothetical measurement on the scatterer and the emitted field. Such studies may guide efforts to optimize probe quantum states and strategies, and we suggest this as a promising avenue for future exploration. 
	
	\begin{acknowledgements}
	This  work  is  supported  by the Danish National Research Foundation through the Center of Excellence “CCQ” (Grant agreement  No.  DNRF156);  and  the  European  QuantERA grant C’MON-QSENS!, by Innovation Fund Denmark GrantNo.  9085-00002B.
	\end{acknowledgements}


	\bibliographystyle{unsrt}

\end{document}